\documentclass[lettersize,journal]{IEEEtran}
\usepackage{amsmath,amsfonts}
\usepackage{tikz}

\usepackage{algorithmic}
\usepackage{algorithm}
\usepackage{array}
\usepackage[caption=false,font=normalsize,labelfont=sf,textfont=sf]{subfig}
\usepackage{textcomp}
\usepackage{stfloats}
\usepackage{url}
\usepackage{verbatim}
\usepackage{graphicx}
\usepackage{cite}
\usepackage{booktabs}
\usepackage{caption}
\usepackage{titlesec}
\usepackage{balance}

\usetikzlibrary{arrows.meta, positioning, calc, fit, backgrounds, shapes.geometric}

\definecolor{mainblue}{RGB}{0, 83, 156}
\definecolor{accentorange}{RGB}{230, 126, 34}
\definecolor{darkgray}{RGB}{60, 60, 60}
\definecolor{lightblue}{RGB}{220, 235, 252}
\definecolor{lightorange}{RGB}{253, 235, 208}
\definecolor{softviolet}{RGB}{138, 110, 180}
\definecolor{lightviolet}{RGB}{235, 230, 245}
\definecolor{softteal}{RGB}{15, 110, 90}
\definecolor{lightteal}{RGB}{214, 238, 232}

\titleformat{\section}
  {\normalfont\scshape\bfseries\centering\large}   
  {}                                 
  {0em}
  {}
\titlespacing{\section}{0pt}{1em}{0.4em}

\titleformat{\subsection}
  {\normalfont\scshape\bfseries\centering\normalsize}   
  {}                                 
  {0em}
  {}
\titlespacing{\subsection}{0pt}{1em}{0.4em}

\titleformat{\subsubsection}
  {\normalfont\scshape\bfseries\centering\small}   
  {}                                 
  {0em}
  {}
\titlespacing{\subsubsection}{0pt}{1em}{0.4em}
  
\begin{document}






\title{\textsc{RRM-GPT}: A Framework and  Vision for \\Radio Resource Management Foundation Models}


\author{
\IEEEauthorblockN{
Ahmed Aboulfotouh\IEEEauthorrefmark{1},
Akram Bin Sediq\IEEEauthorrefmark{2},
Koosha Pourtahmasi Roshandeh\IEEEauthorrefmark{2},
Omar Mashaal\IEEEauthorrefmark{1},\\
Ahmad M. Nagib\IEEEauthorrefmark{2},
Jale Sadreddini\IEEEauthorrefmark{2},
and Hatem Abou-Zeid\IEEEauthorrefmark{1}
}\\
\IEEEauthorblockA{
\IEEEauthorrefmark{1}Department of Electrical and Software Engineering,
University of Calgary, Canada
}\\
\IEEEauthorblockA{
\IEEEauthorrefmark{2}Ericsson Canada Inc., Canada
}
}
\maketitle

\begin{abstract}
Learning-based models for radio resource management (RRM) are typically built for a single function and deployment, so each new setting repeats the development pipeline. RRM decisions, however, share a common structure: each is assembled from interdependent fields, defined by the standard, whose values are selected in view of the network state. We propose RRM-GPT, an autoregressive framework for RRM foundation models that generate these decisions as a language model generates text. An encoder maps heterogeneous network observations into a common token representation, and a decoder emits the decision one field at a time, each conditioned on the network state and the fields already committed. Pretraining on unannotated network logs teaches the model what makes a decision valid and how controllers choose among valid decisions; post-training then adapts it to deployment-specific operator objectives through imitation or reinforcement learning. The framework targets two forms of reuse: a function-specific model reused across deployments, and a model shared across RRM functions that generates their interdependent decisions as one sequence. In a 5G New Radio (NR) case study, we demonstrate that a single model generates complete scheduling grants spanning user selection, timing, link adaptation, resource allocation, and control signaling. The model captures dependencies among grant fields and transfers learned behavior to an unseen scenario without adaptation.
\end{abstract}

\begin{IEEEkeywords}
Radio resource management, autoregressive modeling, wireless foundation models, AI-native RAN, 5G NR scheduling.
\end{IEEEkeywords}

\section{Introduction}
\label{sec:intro}

Every fraction of a millisecond, a cell decides which users to serve, on which spectrum resources, and at what rate. At slower timescales, neighboring cells coordinate spectrum use, users move between cells or carriers, and traffic is redistributed to balance load. Together, these decisions constitute radio resource management (RRM)~\cite{5g_hetnets}. Cellular systems organize them into specialized functions for scheduling, link adaptation, interference coordination, handover, and load management, each with its inputs, constraints, algorithms, and timescale~\cite{3gpp38300}.

With growing interest in AI-native resource management, learning-based methods have been applied successfully to individual RRM functions, such as scheduling~\cite{leasch} and link adaptation~\cite{drllinkadaptation}. Such models are typically specialized to a particular RRM function and deployment scenario, with inputs, outputs, and training objectives tailored to that setting~\cite{sun18, ai6gmac}. Developing a model for a different function or deployment requires revisiting the engineering pipeline: assembling representative data, designing the objective, training, and validating the model. This specialization limits model reuse and increases the engineering effort required to accommodate changes in network configuration, traffic conditions, and operator objectives. Developing models independently for each function also makes it harder to capture dependencies among RRM decisions within a shared context.

\begin{figure*}[t]
    \centering
    \includegraphics[width=\linewidth]{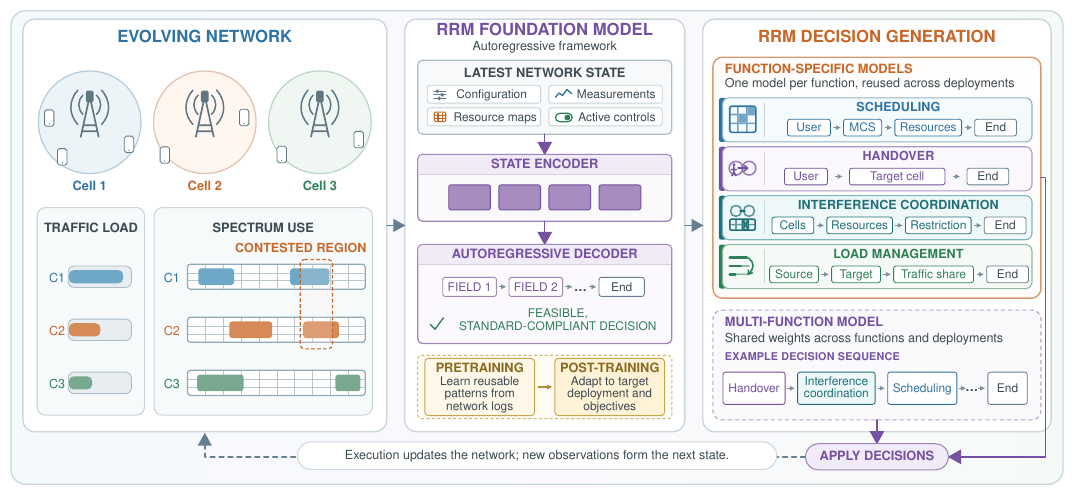}
    \caption{Overview of the proposed autoregressive framework for RRM. Network state is encoded, and feasible, standard-compliant decisions are generated field by field. The formulation supports function-specific models with separate weights and, at its broadest scope, a shared model across functions. Models are pretrained on network logs and post-trained for a deployment objective.}
    \label{fig:framework}
\end{figure*}

To address these limitations, we propose an autoregressive framework, illustrated in Fig.~\ref{fig:framework}, for building RRM foundation models: models pretrained on network logs for reuse across network deployments and RRM functions. The figure distinguishes two scopes of reuse. In the \emph{function-specific} setting, a model supports one RRM function across deployments; a scheduling foundation model, for example, could be adapted to different cell configurations, environments, and traffic conditions. The \emph{multi-function} setting represents a broader ambition: sharing a model across RRM functions so that their decisions can be generated within a common context. Autoregressive Transformer models have been explored for generating scheduling decisions, with a dedicated model for each scheduling subtask~\cite{rrmtransformerpatent}. Prior work on wireless foundation models has primarily explored representation learning from physical-layer signals and channels~\cite{wirelessgpt,wavesfm, liu2025wifo}. The framework presented here provides a common formulation for structured RRM decision generation, with model reuse across deployments and functions as a central objective.


The framework builds on a structural observation: despite their different roles, RRM functions construct decisions in a common form. Each decision is assembled from fields whose values are selected in view of network observations, subject to standards, configuration, and available resources. As illustrated on the right of Fig.~\ref{fig:framework}, a scheduling grant identifies a user, transmission parameters, and radio resources; a handover identifies a user and a target cell; and an interference-coordination decision specifies participating cells, affected time-frequency resources, and restrictions such as muting. 

The fields within a decision are interdependent. In a scheduling grant, channel conditions constrain the rate, the selected rate and queued data determine the resources required, and existing allocations constrain their placement. Dependencies also cross functions: a handover determines which cell subsequently schedules a user, while interference coordination restricts the resources available to affected schedulers. RRM decisions thus exhibit sequential structure at two levels: among fields within a decision, and among decisions across functions.

Our autoregressive framework, RRM-GPT, generates a decision one field at a time, each choice conditioned on the network state and on the fields already committed. Training proceeds in two stages: reusable pretraining and objective-specific post-training. Pretraining uses state-action records collected from operational networks and simulators without manual annotation, from which the model learns how a decision is assembled and how it follows from network conditions. Post-training adapts the model to a target deployment, either by imitating a reference controller or through reinforcement learning to optimize an operator-defined objective expressed in terms of network key performance indicators (KPIs), such as throughput and latency. Our contributions in this article are:
\begin{itemize}
\item We propose an autoregressive framework for generating feasible RRM decisions, with pretraining on network logs and post-training for deployment-specific objectives.
\item We introduce a representation of network observations and RRM decisions that accommodates heterogeneous inputs and function-specific fields and constraints.
\item We outline how the framework can be applied to handover, interference coordination, and load management, and discuss the prospects for model reuse across deployments and functions.
\item We demonstrate complete scheduling-grant generation in a 5G New Radio (NR) case study and evaluate generalization to an unseen scenario.
\end{itemize}

\section{Scheduling as Sequence Generation}
\label{sec:scheduling}

\begin{figure*}[t]
    \centering
    \includegraphics[width=\linewidth]{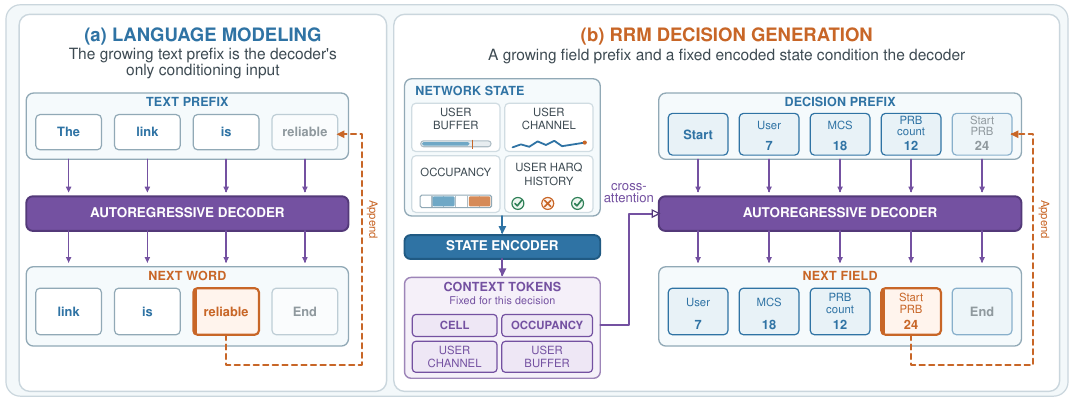}
    \caption{Autoregressive generation in language and RRM. (a) A language model generates each word conditioned on the preceding words. (b) RRM-GPT generates each decision field conditioned on the preceding fields and the encoded network state, which includes resource occupancy, user buffers, channel quality, and hybrid automatic repeat request (HARQ) feedback.}
    \label{fig:llmvsrrm}
\end{figure*}

Consider downlink (DL) scheduling in a 5G NR cell, where time is divided into slots. In each slot, the base station decides which users to serve, how much spectrum each receives, and how its data is encoded. A user cannot know in advance what it has been assigned, so the base station must also send a control message carrying this information and place it where that user is configured to look. A scheduling decision therefore allocates two related kinds of resources: those that carry the data and those that carry the message describing how to receive it \cite{dahlman2020nr}.

The base station makes its choices based on what it observes: how much data each user has waiting and for how long, the channel quality each user reports, and the outcomes of recent transmissions. Consider the example in Fig.~\ref{fig:llmvsrrm}, where the scheduler selects User 7, whose data queue is long. The reported channel quality indicator (CQI) guides the choice of modulation and coding scheme (MCS) 18, which sets how many bits each unit of spectrum carries. That rate and the waiting data determine how much spectrum is needed, twelve physical resource blocks (PRBs), the units in which frequency is assigned. In this example, we assume resource allocation Type 1, which requires allocating consecutive PRBs. The scheduler assigns twelve PRBs one slot ahead, starting at PRB 24. The same channel report shapes the control message: a weaker channel needs more control resources, so it is spread over four control channel elements (CCEs), an aggregation level of 4, beginning at CCE 8, on one of the candidate positions User 7 monitors according to the 3GPP specifications. Together, these choices form one DL grant, and a slot may contain none, one, or several, so how many to issue is itself part of the decision. Because each choice is shaped by those before it, scheduling can be viewed as a sequence of interdependent decisions.

\subsection{From Sentences to Grants}

Although most fields within a grant have only a limited set of possible values, the number of possible grants grows quickly when they are combined. For a single user, choosing the MCS, the slot, the frequency allocation, the control-channel settings, and the remaining fields already yields several million possible grants. A slot may contain several grants, multiplying the possibilities further. Not every combination is valid, either: each grant consumes resources and changes what can be issued next. Representing the entire scheduling decision as one output would therefore require choosing from an enormous set, whose size is not known in advance and whose valid options change as the slot fills.

\textbf{How a language model writes.} Language generation also involves an effectively unbounded set of possible outputs. Rather than selecting a complete sentence from a catalog, an autoregressive model generates it one word at a time, each drawn from a manageable vocabulary and conditioned on the words already written, as shown in Fig.~\ref{fig:llmvsrrm}a. Pretraining on large text corpora teaches the model grammatical and contextual patterns that guide its choices: after “the base station allocates,” for example, words describing resources become plausible continuations. Neither grammatical rules nor likely continuations are explicitly programmed; the model learns both by repeatedly predicting the next word.

\textbf{Generating the grant.} 
A scheduling grant can likewise be generated as a sequence, as illustrated in Fig.~\ref{fig:llmvsrrm}b. The standard defines the grant's fields and their allowed values, much as grammar defines sentence structure. The scheduler's policy selects among valid grants, only a small fraction of which a well-designed scheduler would issue. We generate each grant one field at a time, following a chosen order: the user, the slot, the MCS, the PRB count, the starting PRB, the aggregation level, and the starting CCE. Later fields are conditioned on earlier choices. For example, choosing the MCS before the PRB count allows the model to account for the modulation and coding when determining how many PRBs are needed to serve the queued data. 
The model generates successive grants until the slot's decision is complete. At each step, it selects from a manageable set of values, conditioned on the network state and the fields already generated.

\subsection{Tokens and Vocabularies}

\begin{table*}[t]
\centering
\small
\caption{Correspondence between autoregressive language generation and RRM decision generation. RRM functions differ in their observations, output fields, constraints, and decision scope, while sharing a token-by-token generation process.}
\label{tab:mapping}
\renewcommand{\arraystretch}{1.3}
\setlength{\tabcolsep}{5pt}
\begin{tabular}{@{}
    >{\raggedright\arraybackslash}p{0.12\linewidth}
    >{\raggedright\arraybackslash}p{0.15\linewidth}
    >{\raggedright\arraybackslash}p{0.215\linewidth}
    >{\raggedright\arraybackslash}p{0.20\linewidth}
    >{\raggedright\arraybackslash}p{0.22\linewidth}
@{}}\toprule
 & \textbf{Language model} & \textbf{Scheduling} & \textbf{Interference coordination} & \textbf{Handover} \\
\midrule

\textbf{Input} &
The prompt and text already written &
Per-user queues and channel reports, resource occupancy, cell configuration, and previously generated fields &
Per-cell load, interference between neighbors, and previously generated fields  &
User reports of signal strength for serving and neighboring cells, cell loads, and previously generated fields  \\

\textbf{Token} &
A word or part of a word &
A value for one grant field &
A value for one muting-decision field &
A value for one handover field \\

\textbf{Vocabulary} &
Words and sub-words of a text corpus &
Users, MCS indices, PRB counts and positions, control resources &
Cell identifiers, resource ranges, transmit or mute indicators &
User, cell and carrier identifiers \\

\textbf{Complete sequence} &
A generated text sequence &
All grants issued in one slot &
A muting pattern for one coordination period &
Handovers issued in one decision step \\

\textbf{Example} &
The afternoon sun warmed the garden &
User 7 $\cdot$ MCS 18 $\cdot$ 12 PRBs $\cdot$ from PRB 24 &
Cell 2 $\cdot$ PRBs 30--45 $\cdot$ mute &
User 7 $\cdot$ to Cell 3 \\

\textbf{Rules} &
Grammar and semantics of the language &
3GPP compliance: allocations must not conflict, and control must fall in the user's search space &
Muting must agree between neighbors and fit each cell's resource budget &
The target must be a neighbor the user can measure and is configured to use \\

\textbf{End-of-sequence} &
An end-of-text symbol &
No further grant to issue &
Every cell in the cluster assigned &
No further handover to issue \\

\bottomrule
\end{tabular}
\end{table*}

In language modeling, a \emph{token} is the unit the model emits, such as a word or part of one, and the \emph{vocabulary} is the set of tokens it may choose from, constructed from a text corpus. Both have direct counterparts in RRM-GPT. Each output token specifies the value of one decision field: for example, when generating the MCS field, the model selects an index from a configured MCS table. That set of indices is the field's vocabulary. Unlike a text vocabulary, these choices come from the 3GPP standards and network configuration. 

We set out this correspondence for relevant RRM functions in Table \ref{tab:mapping}. A scheduling sequence specifies the grants issued within a slot, an interference-coordination sequence specifies a muting pattern, and a handover sequence specifies which users move and to which target cells. A sequence may contain zero or more such actions, each spanning multiple tokens, with an end-of-sequence token marking completion of the decision step. The same generation principle therefore carries across different RRM functions and decisions.

In contrast to text, the input to RRM-GPT has no shared, predefined vocabulary and varies with the use case. The inputs include the cell's configuration, per-user reports such as the CQI and queued data, and the cell's resource occupancy as a two-dimensional map. These observations are heterogeneous, and none arrives as a token. We propose encoding each input according to its structure: categorical values through a learned dictionary, continuous values by their position within learned intervals to retain magnitude information, and the occupancy map in patches to preserve its layout. 
Related features are then grouped and projected into a common embedding space to form input tokens. For example, a user's CQI and SINR estimate could form a channel token, while queue lengths and waiting times could form a demand token. The model thus receives a uniform sequence of tokens describing each user, the cell, and its resources.

\subsection{Encoder--Decoder Architecture}

RRM decisions depend on relationships among users and shared resources. A user's queue indicates its demand, but deciding whether and how to serve it also requires considering its channel conditions, competing users, and resource availability. We choose an encoder--decoder architecture to separate joint processing of the network state from sequential decision generation, as illustrated in Fig.~\ref{fig:llmvsrrm}b.

The encoder processes the input tokens jointly through self-attention, allowing each token to incorporate information from the rest of the state. For example, a user's demand token can draw on its channel information and the resource-occupancy tokens, providing the decoder with a representation of that demand in its network context. The encoder preserves a sequence of such representations, which the decoder can attend to selectively as it generates different fields.

The autoregressive decoder generates the decision one field at a time, drawing on two sources of context: the encoded network state and the growing sequence of fields already chosen~\cite{vaswaniattention}. The state is encoded once per decision step and reused throughout generation, so the network state is processed only once however many fields the decision contains.

Autoregressive decoding expresses the probability of a complete decision as a product of conditional probabilities, one per field. This decomposition is exact and makes no independence assumption among fields: the decoder selects a PRB count with knowledge of the user’s MCS and places a second grant with knowledge of the resources assigned to the first. Sequential generation mirrors how a slot is filled: each commitment narrows the options available to the next, allowing constraints to be enforced as the decision is constructed. Each field is therefore chosen with the full network state in view and every earlier choice already fixed.

\subsection{Learning from Network Logs}

Pretraining uses recorded network observations paired with the RRM decisions taken in response. For scheduling, each record contains the cell and user states, and the grants issued in a slot. Logs of this kind already exist in volume: vendors and operators record scheduling decisions continuously during normal operation, and simulators can generate them at scale. As illustrated in Fig.~\ref{fig:training}, the model is given the network state and the preceding fields of the recorded decision and trained to predict the next field, following the same next-token prediction objective used in language modeling.

Logged RRM decisions provide two kinds of information: the rules that determine whether a decision is valid, and the controller's behavior when choosing among valid decisions. For example, valid scheduling records avoid conflicting resource allocations and place control information on resources the scheduled user is configured to monitor under 3GPP specifications. The choice of which users receive grants and which MCS values they use reflects the controller's policy. Predicting logged fields allows the model to learn both validity rules and decision-making behavior without explicitly implementing the controller's algorithms, much as predicting text teaches a language model both grammar and semantics. 

The policy learned through imitation is limited by the behavior represented in the logs. Objective-driven post-training can subsequently refine that policy, with the aim of improving network performance beyond that of the reference controller.

\subsection{Guaranteeing Feasibility}

\begin{figure*}[t]
    \centering
    \includegraphics[width=\linewidth]{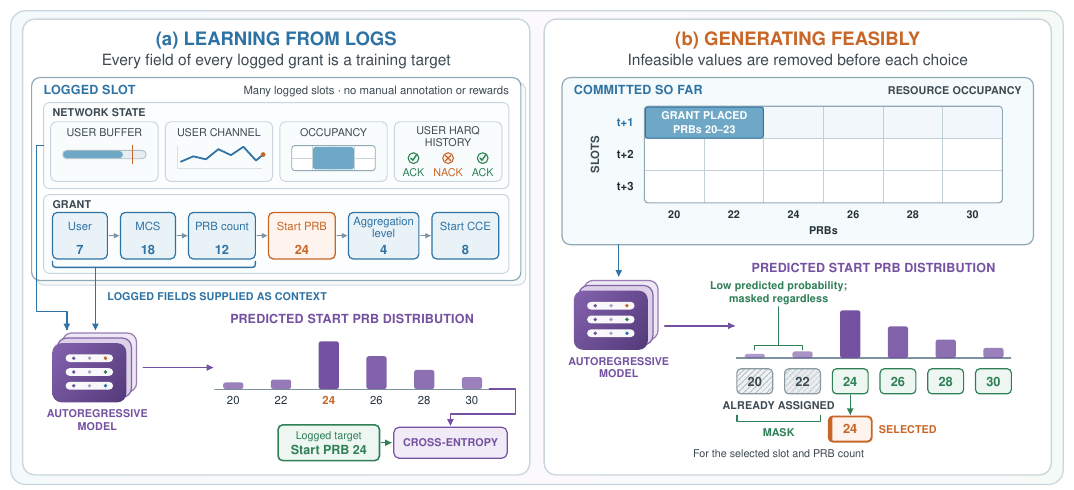}
    \caption{Training and generation at the same step, the starting PRB of User 7's grant. (a) Every field of a logged grant is a supervised target, predicted from the state and the fields before it and scored against the logged value. (b) At generation, values conflicting with what the slot has committed are removed first; the model already puts almost no weight on them.}
    \label{fig:training}
\end{figure*}

The rules learned from logs capture statistical regularities but do not provide hard guarantees. While in language this may produce unconventional content, in RRM an invalid grant or control information placed outside a user's monitored search space can directly disrupt network operation.

Moreover, feasibility is conditional: the validity of each value depends on previously selected fields and resources already committed within the slot. Earlier allocations restrict the resources available to subsequent users, and the remaining control-channel capacity decreases as grants are issued. Because each grant is generated field by field, these constraints can be enforced during decoding, as shown in Fig.~\ref{fig:training}b. At each step, values that would violate the current constraints are masked before selection, ensuring that the model chooses only from the remaining feasible values. A lightweight state tracker records the selected field values and resource commitments, updating them after each generated field so that the mask can determine which values remain feasible at the next step. 

Masking is applied only during inference to ensure feasibility. A model trained on sufficient logs largely respects these rules on its own, so masking rarely changes the outcome. Even rare violations can disrupt network operation, making such lightweight guarantees essential for practical deployment.

\section{Beyond Scheduling: Autoregressive Modeling of Joint RRM}
\label{sec:joint_rrm}

The scheduling formulation above generates all grants within a slot as one decision sequence. It takes the user's serving cell and any resource restrictions imposed by inter-cell coordination as given. These conditions are themselves shaped by other RRM functions, including handover, interference coordination, and load management.

Extending to functions beyond scheduling does not require a new generation mechanism. The framework accommodates these functions naturally by moving the boundary of what the model treats as a decision. As shown in Fig.~\ref{fig:framework}, a shared model can generate load-management, handover, interference-coordination, and scheduling decisions autoregressively as an interdependent sequence. Decisions made by other RRM functions, previously treated as inputs to scheduling, can now be generated by the shared model, allowing decisions across functions to condition on one another.

\subsection{A Unified Generative Architecture}

Sharing a model does not require all RRM functions to act at every step. Each function retains its own decision structure and control timescale and is invoked when its control opportunity arises. In a slot-based implementation, most steps would contain scheduling grants alone, while some would also include handover, interference-coordination, or load-management decisions. Generation uses only the observations and prior decisions available at the relevant time and network scope. Joint modeling therefore does not assume access to a complete, synchronized network state.

When several functions are invoked together, the generation order follows their dependencies. An interference-coordination decision precedes the grants whose resources it constrains, while a handover decision conditions scheduling for the affected user. Slower decisions need not be regenerated at every slot: their continuing effects remain part of the network state and condition faster decisions until they are revised. The functions share the model and its decision context while retaining their own decision structures and control timescales.

The value of this architecture lies in modeling interactions among RRM functions. An overloaded cell may be relieved by moving its users to a neighbor, by coordinating resources with neighboring cells, or by changing which users are scheduled. Handover, load management, and interference coordination each change the problem presented to the others. Their decisions can complement or substitute for one another, making it useful to model them jointly. A unified model captures these dependencies within one policy. Autoregressive generation breaks this broader joint decision into a sequence of smaller choices, each conditioned on those already committed.

\subsection{Post-Training and Specialization}

Pretraining provides a common initialization, but the behavior needed in a particular deployment may differ from that represented in the original logs. During post-training, the pretrained model retains the same state representation and autoregressive output structure, while its weights are adapted using deployment-specific data or feedback. This may update the full model or only a small set of specialization parameters attached to the shared backbone.

\textbf{Behavioral cloning.}
The pretrained model can be fine-tuned to imitate a reference controller's logged decisions, using the same prediction objective as pretraining without an explicit reward. This offers a way to approximate a computationally expensive controller: high-quality decisions generated offline can train a model intended for real-time operation. Pretraining on logs from different controllers exposes the model to multiple scheduling approaches; behavioral cloning then specializes it to a target deployment, potentially requiring fewer demonstrations than training from scratch.

\textbf{Reinforcement learning.}
Alternatively, the model can be optimized directly for desired network outcomes. The model generates RRM decisions, and their effects on network performance are evaluated using an operator-defined objective. Reinforcement learning uses this feedback to adjust the probabilities of the generated decisions to increase the expected long-term return, similar to RL-based post-training of language models~\cite{deepseek}.
Throughput, latency, fairness, and energy efficiency can inform the reward, with their relative importance reflecting deployment priorities. Multi-objective reinforcement learning offers a further extension: conditioning the policy on these priorities and training across different trade-offs could allow one model to support changing operator preferences without retraining for each setting~\cite{demirel2026intentsactionsagenticai}.

In each case, post-training builds on the relationships between network observations and decisions learned during pretraining. Slow convergence and the training required to learn effective RRM policies motivate methods that accelerate learning through knowledge transfer~\cite{nagibdrl}. Reusing a pretrained model offers a way to address this challenge, potentially reducing the deployment-specific data or interaction needed to adapt to a target controller or objective. 

The following case study instantiates the framework for 5G NR scheduling. It evaluates whether the proposed autoregressive model can learn and generate coupled scheduling decisions from network logs.

\section{Case Study: 5G NR Scheduling}
\label{sec:results}

\subsection{Experimental Setup}

We generate scheduling logs using a system-level 5G NR simulator with a reference scheduler. The simulated urban macro cell operates at 3.5~GHz over 40~MHz, with 106 PRBs, numerology $\mu=1$, and a time-division duplexing pattern of three downlink (DL) slots followed by one uplink (UL) slot. Users request back-to-back 100~KB FTP files. The training set contains approximately 6.5 million grants from runs lasting up to 60~s, with 5, 10, or 20 active users moving at 50~km/h within a cell of radius 1~km. We evaluate on 637,000 grants from held-out runs of the same scenario. To assess generalization, we also test on 158,000 grants from a separate scenario with static users in a cell of radius 167~m and the same active-user counts. This scenario is excluded entirely from training.

\begin{figure*}[h!]
    \centering
    \includegraphics[width=0.86\textwidth]{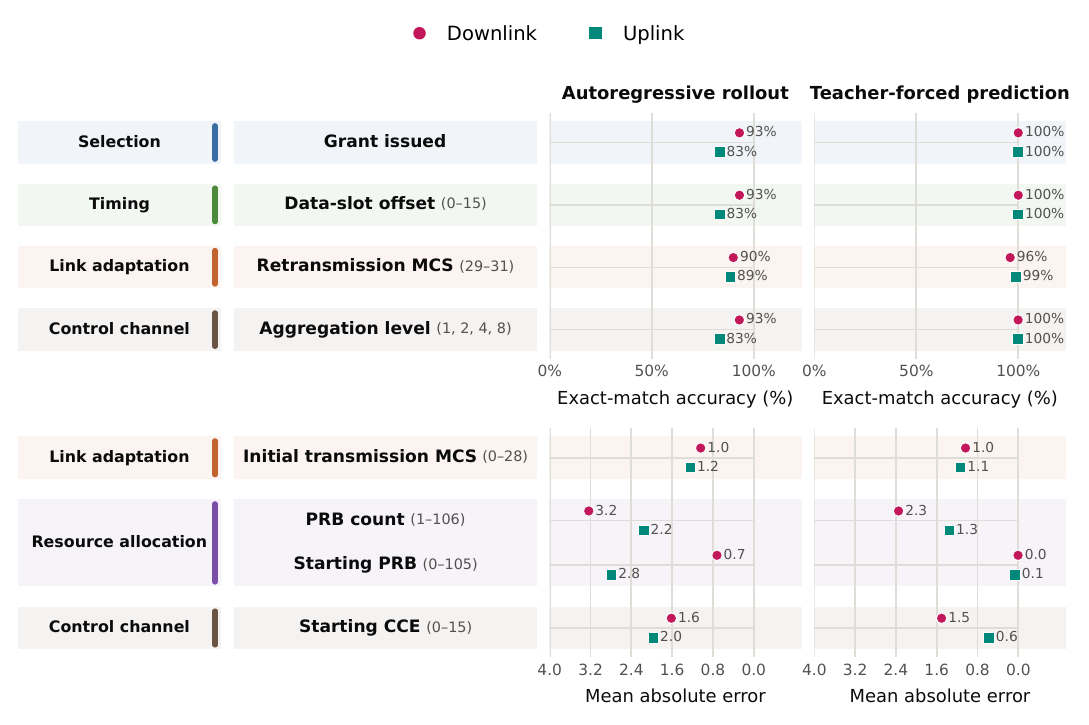}
    \caption{Scheduling-field prediction on the mobile test set under autoregressive rollout and teacher forcing: exact-match accuracy (top) and mean absolute error in field units (bottom). Parentheses give each field's values or range.}
    \label{fig:field_scores}
\end{figure*}
\begin{figure*}[h!]
    \centering
    \includegraphics[width=0.86\textwidth]{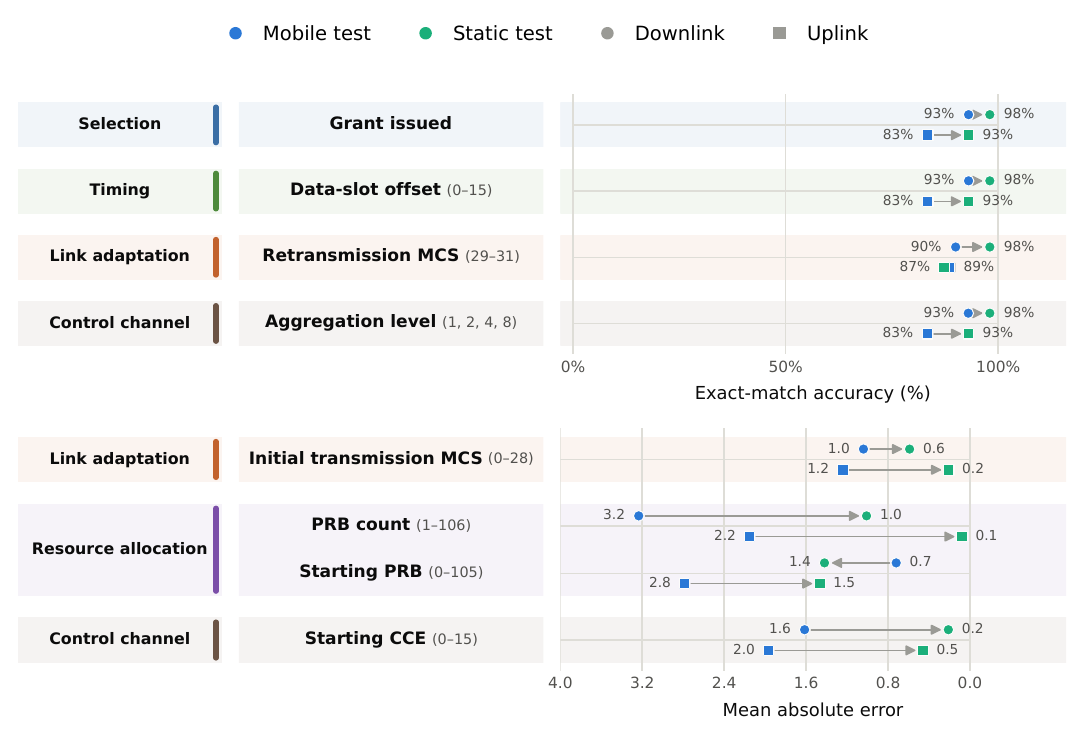}
    \caption{Generalization of the mobile-trained model to the held-out static scenario without adaptation, under autoregressive rollout: exact-match accuracy (top) and mean absolute error in field units (bottom). Arrows show the change from mobile to static testing.}
    \label{fig:generalization}
\end{figure*}

At each slot, the model generates zero or more grants. Each grant specifies the selected user, data-slot offset, MCS, PRB count, starting PRB, control-channel aggregation level, and starting CCE, in that order. These fields span user selection, timing, link adaptation, resource allocation, and control signaling. An end-of-sequence token marks completion of the slot decision. Grants are issued for both DL and UL transmissions, and each schedules either an initial transmission or a retransmission.

The model receives the cell configuration, a time-frequency occupancy grid, and per-user CQI, buffered data, HARQ feedback, and control-channel search space. The encoder and decoder each comprise two Transformer layers with an embedding dimension of 256 and four attention heads per layer, yielding approximately 4.3 million parameters. We train for 50 epochs using AdamW with a batch size of 128.

\subsection{Numerical Results}

We compare autoregressive rollout with teacher-forced prediction (Fig.~\ref{fig:field_scores}). During rollout, each prediction is conditioned on the model’s previously generated fields. Under teacher forcing, the model instead receives the preceding ground-truth fields and predicts the next field, which is scored against its logged value. The logged value is then used to predict subsequent fields. This comparison helps distinguish field-prediction difficulty from the effects of errors in earlier choices. Generated grants are matched to logged grants by user, so grant ordering does not affect the scores. We report mean absolute error (MAE) for fields where the size of a mismatch is informative, such as MCS and resource positions, and exact-match accuracy for the remaining fields. Missing grants count as errors in exact-match accuracy and are excluded from MAE; grant-issued accuracy separately measures whether a grant was generated for the logged user.

The results show that the model learns dependencies among grant fields and closely reproduces much of the reference scheduler’s behavior during rollout. Timing and aggregation level match the grant-issued accuracy of 93\% in DL and 83\% in UL, so their remaining errors stem from missing grants rather than from the fields themselves. Initial-transmission MCS deviates from the logged choice by about one index on average, with MAEs of 1.0 and 1.2 for DL and UL. Under teacher forcing, all accuracy-scored fields exceed 95\% and UL starting-PRB MAE falls from 2.8 to 0.1 indices. The gains are largest for starting PRB, which depends on the grant’s PRB count and on the resources already assigned to users scheduled earlier in the slot. Given the correct preceding fields, the model thus reproduces most choices almost exactly, suggesting that error propagation accounts for much of the rollout mismatch.

We next evaluate the model on the held-out static scenario without adaptation (Fig.~\ref{fig:generalization}). Performance improves for most fields under the change in mobility and cell radius: grant issuance, timing, and aggregation level all gain accuracy, and MAE falls for initial-transmission MCS, most markedly in UL, from 1.2 to 0.2 indices, and for UL starting PRB. Static users and a smaller cell yield more stable channel conditions, which may ease prediction; the gains nonetheless show that the model has not memorized the training scenario. Only DL starting-PRB MAE rises, from 0.7 to 1.4 PRB indices. The same model thus reproduces substantial aspects of the reference scheduler’s behavior in a scenario excluded from training, supporting reuse across deployments.

Together, these results support the function-specific scope of the proposed framework: one autoregressive model generates the fields of scheduling grants and transfers learned behavior to an unseen deployment scenario. They provide an initial basis for model reuse, while broader generalization, the gains from post-training, and network-level performance remain open for future evaluation.

\section{Conclusion and Open Problems}
\label{sec:conclusion}

RRM decisions retain a common structure across deployments, even as cell configurations, propagation conditions, traffic demands, and operator objectives change. We proposed an autoregressive framework that exploits this structure to learn RRM decisions from network logs, with post-training as the route to deployment-specific adaptation. Our 5G NR case study demonstrates complete scheduling-grant generation and  generalization to an unseen scenario without adaptation. Establishing reliable adaptation across a wider range of deployments and objectives is the next step toward RRM foundation models. Sharing models across RRM functions represents a broader ambition, extending reuse to interdependent decisions taken at different scopes and timescales. Pursuing these directions requires addressing several open research problems.

\textbf{Representation and transfer across systems.}
The framework expresses each decision as an ordered set of fields, but which fields to use, their granularity, and order remain design questions. A representation suited to one standard release may not transfer to the next, and the vocabularies describing 5G NR grants differ from those of Wi-Fi or satellite systems. An open question is whether a shared representation can be defined that is general enough to transfer across systems yet specific enough to generate valid decisions, allowing a model pretrained on one network to provide a useful starting point for another.

\textbf{Network logs and data coverage.}
Pretraining depends on the coverage and detail of network logs. Models intended for reuse across deployments require records spanning different configurations, channel conditions, traffic demands, and controller policies. Extending learning across RRM functions also requires traces that preserve the timing and dependencies between their decisions, together with the relevant network observations. Establishing what to record, at what granularity, and how to combine logs from different vendors and deployments is therefore a central research challenge.

\textbf{Complexity, latency, and deployment.}
Decisions taken every slot impose tight inference budgets, while decisions and inputs processed every few minutes allow greater computational complexity. Architectures should therefore avoid repeating expensive computation at fast timescales. Where models run and how their instances coordinate are also open questions. A single pretrained model may be deployed at multiple points in the network, each operating on information available to its local control loop. Maintaining consistency among decisions made from these different local observations remains an important challenge.

\textbf{Expressing and pursuing operator intent.}
Post-training requires a well-defined objective, yet operators may not always be able to specify one explicitly. Some goals are directly quantifiable, while others are more naturally expressed through relative preferences, and many vary with time, deployment conditions, and commercial priorities. This raises two key questions: how operational intent should be expressed and how to adapt a model to that intent without unsafe exploration in a live network. Supporting diverse objectives in a single model through lightweight specialization is a related challenge.

\bibliography{library}
\bibliographystyle{ieeetr}
\balance

\end{document}